\documentclass[3p]{elsarticle}

\usepackage{lineno,hyperref}
\usepackage{amsmath,amssymb}
\usepackage{subcaption}
\modulolinenumbers[1]

\journal{Journal of Computational Physics}

\begin{document}

\begin{frontmatter}

\title{A mass-energy-conserving discontinuous Galerkin scheme for the isotropic multispecies Rosenbluth--Fokker--Planck equation}


\author{Takashi Shiroto\corref{mycorrespondingauthor}}
\ead{shiroto.takashi@qst.go.jp}
\cortext[mycorrespondingauthor]{Corresponding author}

\author{Akinobu Matsuyama}
\author[naka]{Nobuyuki Aiba}
\author{Masatoshi Yagi}

\address{Rokkasho Fusion Institute, National Institutes for Quantum and Radiological Science and Technology,
2-166 Oaza-Obuchi-Aza-Omotedate, Rokkasho, Aomori 039-3212, Japan}
\address[naka]{Naka Fusion Institute, National Institutes for Quantum and Radiological Science and Technology,
801-1 Mukoyama, Naka, Ibaraki 311-0193, Japan}

\begin{abstract}
Structure-preserving discretization of the Rosenbluth--Fokker--Planck equation is still an open question
especially for unlike-particle collision.
In this paper, a mass-energy-conserving isotropic Rosenbluth--Fokker--Planck scheme is introduced.
The structure related to the energy conservation is skew-symmetry in mathematical sense,
and the action--reaction law in physical sense.
A thermal relaxation term is obtained by using integration-by-parts
on a volume integral of the energy moment equation, so the discontinuous Galerkin method is selected
to preserve the skew-symmetry.
The discontinuous Galerkin method enables ones to introduce the nonlinear upwind flux without violating the conservation laws.
Some experiments show that the conservative scheme maintains the mass-energy-conservation only with round-off errors,
and analytic equilibria are reproduced only with truncation errors of its formal accuracy.
\end{abstract}


\end{frontmatter}


\section{Introduction}\label{sec:1}
Rosenbluth--Fokker--Planck (RFP) equation \cite{Rosenbluth1957} describes stochastic relaxation
of the distriution function through small-angle scattering of Coulomb collision.
The RFP equation is composed of a nonlinear Fokker--Planck equation and Rosenbluth potential equations.
The nonlinear Fokker--Planck equation is a convection--diffusion partial-differential-equation (PDE) described as follows:

\begin{align}
\frac{\partial f_\mathrm{s}}{\partial t}=\Gamma_{\mathrm{s/s'}}
\frac{\partial}{\partial\mathbf{u}}\cdot\left[\frac{\partial^2 G_\mathrm{s'}}{\partial\mathbf{u}\partial\mathbf{u}}
\cdot\frac{\partial f_\mathrm{s}}{\partial\mathbf{u}}
-\frac{m_\mathrm{s}}{m_\mathrm{s'}}\frac{\partial H_\mathrm{s'}}{\partial\mathbf{u}}f_\mathrm{s}\right],\label{eq:1.1}\\
\Gamma_{\mathrm{s/s'}}=\frac{2\pi Z_\mathrm{s}^2Z_\mathrm{s'}^2e^4\ln\Lambda_\mathrm{s/s'}}{m_\mathrm{s}^2},\label{eq:1.2}
\end{align}
where $f$ is the distribution function, $m$ is the mass, $\mathbf{u}$ is the momentum per unit mass,
$Z$ is the mean charge, $e$ is the elementary charge, $\ln\Lambda$ is the Coulomb logarithm,
and $(\mathrm{s,s'})$ are the labels of particle species.
Equation~(\ref{eq:1.2}) satisfies the following relation, which is essential for the momentum and energy conservation:

\begin{align}
m_\mathrm{s}^2\Gamma_\mathrm{s/s'}=m_\mathrm{s'}^2\Gamma_\mathrm{s'/s}.\label{eq:1.2-1}
\end{align}
The scalar potentials $H$ and $G$ are obtained by the following Poisson equations:

\begin{align}
\frac{\partial}{\partial\mathbf{u}}\cdot\frac{\partial H_\mathrm{s'}}{\partial\mathbf{u}}=-8\pi f_\mathrm{s'},\quad
\frac{\partial}{\partial\mathbf{u}}\cdot\frac{\partial G_\mathrm{s'}}{\partial\mathbf{u}}=H_\mathrm{s'}.\label{eq:1.3}
\end{align}
These scalar potentials determine the transport coefficients appearing in Eq.~(\ref{eq:1.1}).
The RFP equation is equivalent to the non-relativistic Landau--Fokker--Planck (LFP) equation \cite{Landau--Lifshitz}
described as follows:

\begin{align}
\frac{\partial f_\mathrm{s}}{\partial t}=\Gamma_\mathrm{s/s'}\frac{\partial}{\partial\mathbf{u}}\cdot
\iiint\mathsf{U}(\mathbf{u,u'})\cdot\left(f_\mathrm{s'}\frac{\partial f_\mathrm{s}}{\partial\mathbf{u}}
-\frac{m_\mathrm{s}}{m_\mathrm{s'}}f_\mathrm{s}\frac{\partial f_\mathrm{s'}}{\partial\mathbf{u'}}\right)
\mathrm{d}\mathbf{u'},\label{eq:1.4}\\
\mathsf{U}(\mathbf{u,u'})=\frac{|\mathbf{u-u'}|^2\mathsf{I}-\mathbf{(u-u')\otimes(u-u')}}{|\mathbf{u-u'}|^3},\label{eq:1.5}
\end{align}
where $\mathsf{I}$ is the unit tensor.

Although the RFP equation (\ref{eq:1.1}) is identical to the LFP equation (\ref{eq:1.4}),
they have clearly different computational aspects.
The RFP equation is a system of PDEs so computational cost per time step
can be $\mathcal{O}(N)$ if the most efficient solver, e.g. multigrid method \cite{Brandt1977}, is employed.
Here, $N$ is the number of unknowns.
The RFP equation calculates potential fields from the distribution function,
and each particle interacts with those;
individual binary collisions are masked behind the Rosenbluth potentials.
In contrast, the LFP equation is an integro-differential equation,
so computational cost per time step is $\mathcal{O}(N^2)$;
the LFP equation counts up all of binary collisions,
and suffers ``the curse of dimensionality'' which comes from multiple integrals.
Fast multipole method (FMM) is one of the candidates to reduce the computational cost
of multiple integrals dramatically \cite{Greengard1997}.
The FMM is originally proposed as $\mathcal{O}(N)$ numerical approximation for the $N$-body problem,
and it is also employed to reduce complexity of the multiple integrals.
Some fast conservative schemes with complexity of $\mathcal{O}(N)$ have been proposed,
but it seems that accuracy of the fast algorithms such as FMM is degraded \cite{Buet1997,Lemou2000}.
Another approach to reduce the cost of multiple integrals is quantum computing.
It was reported that a quantum algorithm can perform multiple integrals with
complexity of $\mathcal{O}(1/\varepsilon)$ although classical computing requires
$\mathcal{O}(1/\varepsilon^2)$ operations, where $\varepsilon$ represents the error of numerical solutions \cite{Abrams1999}.
Recently, a promising Vlasov--Maxwell algorithm based on quantum computing was published \cite{Engel2019},
and probably classical algorithms will be driven out by quantum algorithms in the future.
However, the fastest quantum computer is much slower than classical computers as of 2020,
so we should continue to develop conservative Fokker--Planck schemes within classical computing.

The conservation laws of mass, momentum and energy are necessary conditions for the thermal equilibrium.
Hence, many conservative schemes have been discussed for the nonlinear Fokker--Planck equations.
A conservative scheme for the nonlinear LFP equation was developed for the isotropic
distribution function at the dawn \cite{Bobilev1980},
and further developments of multidimensional \cite{Pekker1984,Hirvijoki2017}
and entropic schemes \cite{Berezin1987,Degond1994,Buet1998,Buet2006,Buet2007}
have been carried out later.
Recently, some conservative schemes for the relativistic LFP equation were
developed in the finite-difference \cite{Shiroto2019-1}
and finite-element \cite{Hirvijoki2019} approaches.
The kernel Eq.~(\ref{eq:1.5}) has symmetries described as follows:

\begin{align}
\mathsf{U}(\mathbf{u,u'})=\mathsf{U}(\mathbf{u',u}),\label{eq:1.6}\\
\mathsf{U}(\mathbf{u,u'})\cdot\mathbf{u}=\mathsf{U}(\mathbf{u',u})\cdot\mathbf{u'}.\label{eq:1.7}
\end{align}
These symmetries are related to the conservation laws of momentum and energy respectively,
so most of the conservative LFP schemes have been developed by the ``structure-preserving'' strategy.
However, development of the structure-preserving RFP schemes is more difficult
than that of the structure-preserving LFP schemes.
An energy-conserving structure-preserving scheme was developed for the
single-species RFP equation \cite{Chacon2000-1,Chacon2000-2}.
The scheme is based on a tensor formalism of the RFP equation,
which is similar to the Maxwell stress tensor in electromagnetism.
The Maxwell stress tensor consists the stress-energy tensor,
so the tensor formalism has a deep relation with the conservation laws.
However, Ref.~\cite{Chacon2000-1} points out two remaining issues,
i.e., boundary condition and unlike-particle collision.
The scheme is based on the finite-difference method,
and momentum-energy-conserving boundary conditions are difficult to be applied.
In addition, the tensor formalism was derived only for like-particle collision
because the energy equation for unlike-particle collision cannot be described in the conservative formulation.
Another approach is to introduce ``nonlinear constraints'' which artificially modify
the numerical flux to enforce the conservation laws \cite{Taitano2015,Taitano2016,Taitano2017,Taitano2018,Daniel2020}.
However, there are countless candidates for the nonlinear constraints
because they do not depend on detailed discussions on mathematical structure of the RFP equation.

Here we report a mass-energy-conserving scheme for the RFP equation.
As a proof-of-principle, we focus on isotropic geometry in this paper.
The discontinuous Galerkin (DG) method is chosen since
the conservation laws are expressed as weak formulations of the Fokker--Planck equation,
and the conservative boundary condition is easy to be implemented as the numerical flux.
The rest of this paper is as follows.
In Sec.~\ref{sec:2}, the structure which should be preserved in unlike-particle collision
is revealed through derivation of the conservation laws.
Actual implementation of the proposed scheme is shown in Sec.~\ref{sec:3}.
Some numerical experiments are performed in Sec.~\ref{sec:4} to verify the computational theory.
Section~\ref{sec:5} is the conclusions of this paper.

\section{Analytic derivation of conservation laws}\label{sec:2}
In the rest of this paper, the distribution function is assumed to be isotropic,
so Eqs.~(\ref{eq:1.1}) and (\ref{eq:1.3}) are expressed as follows:

\begin{align}
\frac{\partial f_\mathrm{s}}{\partial t}=\Gamma_{\mathrm{s/s'}}\frac{1}{u^2}
\frac{\partial}{\partial u}\left(u^2\frac{\partial^2 G_\mathrm{s'}}{\partial u^2}\frac{\partial f_\mathrm{s}}{\partial{u}}
-\frac{m_\mathrm{s}}{m_\mathrm{s'}}u^2\frac{\partial H_\mathrm{s'}}{\partial u}f_\mathrm{s}\right),\label{eq:2.1}\\
\frac{1}{u^2}\frac{\partial}{\partial u}\left(u^2\frac{\partial H_\mathrm{s'}}{\partial u}\right)=-8\pi f_\mathrm{s'},\quad
\frac{1}{u^2}\frac{\partial}{\partial u}\left(u^2\frac{\partial G_\mathrm{s'}}{\partial u}\right)=H_\mathrm{s'},\label{eq:2.2}
\end{align}
where $u=|\mathbf{u}|$ is the norm of momentum per unit mass.
The conservation laws are discussed as weak formulations of Eq.~(\ref{eq:2.1}).
Note that the law of momentum conservation is automatically preserved since
the distribution function is isotropic.
The conservation laws of mass and energy are discussed in the following subsections.

\subsection{The mass conservation}\label{sec:2.1}
The mass conservation is expressed as a zeroth-order moment of Eq.~(\ref{eq:2.1}):

\begin{align}
\int_{u_0}^{u_1} \frac{\partial f_\mathrm{s}}{\partial t}m_\mathrm{s}u^2\mathrm{d}u
=m_\mathrm{s}^2\Gamma_\mathrm{s/s'}\left[\frac{1}{m_\mathrm{s}}u^2\frac{\partial^2 G_\mathrm{s'}}{\partial u^2}
\frac{\partial f_\mathrm{s}}{\partial u}
-\frac{1}{m_\mathrm{s'}}u^2\frac{\partial H_\mathrm{s'}}{\partial u}f_\mathrm{s}\right]_{u_0}^{u_1},\label{eq:2.1.1}
\end{align}
where $u^2$ is the Jacobian of the isotropic coordinate,
and the control volume is a spherical shell whose inner and outer radii are $u_0$ and $u_1$, respectively.
Mass in the control volume only depends on surface integrals,
so the RFP equation naturally satisfies the law of mass conservation.

\subsection{The energy conservation}\label{sec:2.2}
The energy moment equation is expressed as a second-order moment of Eq.~(\ref{eq:2.1}):

\begin{align}
\int_{u_0}^{u_1} \frac{\partial f_\mathrm{s}}{\partial t}m_\mathrm{s}\frac12u^4\mathrm{d}u
&=m_\mathrm{s}^2\Gamma_\mathrm{s/s'}\left[\frac{1}{2m_\mathrm{s}}u^4\frac{\partial^2 G_\mathrm{s'}}{\partial u^2}
\frac{\partial f_\mathrm{s}}{\partial u}
-\frac{1}{2m_\mathrm{s'}}u^4\frac{\partial H_\mathrm{s'}}{\partial u}f_\mathrm{s}\right]_{u_0}^{u_1}\notag\\
&-m_\mathrm{s}^2\Gamma_\mathrm{s/s'}\int_{u_0}^{u_1}u^3\left(\frac{1}{m_\mathrm{s}}\frac{\partial^2 G_\mathrm{s'}}{\partial u^2}
\frac{\partial f_\mathrm{s}}{\partial u}-\frac{1}{m_\mathrm{s'}}
\frac{\partial H_\mathrm{s'}}{\partial u}f_\mathrm{s}\right)\mathrm{d}u.\label{eq:2.2.1}
\end{align}
Usually, Eq.~(\ref{eq:2.2.1}) is used as one of the discretized equations in the DG method,
but the law of energy conservation can be violated numerically due to truncation errors.
What is problem is the relation between the energy moment equation for each species Eq.~(\ref{eq:2.2.1})
and conservation of total energy is unclear.
The energy of each species is relaxed through the interaction between species ``$\mathrm{s}$'' and ``$\mathrm{s'}$,''
so the energy moment equation of each species cannot be described in the conservative form.
Therefore, the law of energy conservation should be discussed as sum of those.

By using integration-by-parts to the volume intgral of Eq.~(\ref{eq:2.2.1}),

\begin{align}
\int_{u_0}^{u_1} \frac{\partial f_\mathrm{s}}{\partial t}m_\mathrm{s}\frac12u^4\mathrm{d}u
&=m_\mathrm{s}^2\Gamma_\mathrm{s/s'}\left[\frac{1}{2m_\mathrm{s}}u^4\frac{\partial^2 G_\mathrm{s'}}{\partial u^2}
\frac{\partial f_\mathrm{s}}{\partial u}
-\frac{1}{2m_\mathrm{s'}}u^4\frac{\partial H_\mathrm{s'}}{\partial u}f_\mathrm{s}\right]_{u_0}^{u_1}
-m_\mathrm{s}^2\Gamma_\mathrm{s/s'}\left[\frac{1}{m_\mathrm{s}}u^3
\frac{\partial^2 G_\mathrm{s'}}{\partial u^2}f_\mathrm{s}\right]_{u_0}^{u_1}\notag\\
&-\frac{m_\mathrm{s}^2\Gamma_\mathrm{s/s'}}{8\pi m_\mathrm{s}}
\left[u^2\frac{\partial}{\partial u}\left(uH_\mathrm{s'}\right)\frac{\partial H_\mathrm{s}}{\partial u}\right]_{u_0}^{u_1}
+m_\mathrm{s}^2\Gamma_\mathrm{s/s'}\int_{u_0}^{u_1}
\left(\frac{u^3f_\mathrm{s}}{m_\mathrm{s'}}\frac{\partial H_\mathrm{s'}}{\partial u}
-\frac{u^3f_\mathrm{s'}}{m_\mathrm{s}}\frac{\partial H_\mathrm{s}}{\partial u}\right)\mathrm{d}u.\label{eq:2.2.3}
\end{align}
The point of this paper is that the thermal relaxation term, which is identical to the volume integral of Eq.~(\ref{eq:2.2.3}),

\begin{align}
R_\mathrm{s/s'}=m_\mathrm{s}^2\Gamma_\mathrm{s/s'}\int_{u_0}^{u_1}
\left(\frac{uf_\mathrm{s}}{m_\mathrm{s'}}\frac{\partial H_\mathrm{s'}}{\partial u}
-\frac{uf_\mathrm{s'}}{m_\mathrm{s}}\frac{\partial H_\mathrm{s}}{\partial u}\right)u^2\mathrm{d}u,\label{eq:2.2.4}
\end{align}
has skew-symmetry between species ``$\mathrm{s}$'' and ``$\mathrm{s'}$.''
Owing to the skew-symmetry and Eq.~(\ref{eq:1.2-1}),
the energy conservation can be derived since $R_\mathrm{s/s'}+R_\mathrm{s'/s}=0$.
The distribution function ``$f$'' and potential ``$H$'' may include truncation errors after discretization,
but such errors behave on Eq.~(\ref{eq:2.2.4}) skew-symmetrically and are cancelled out exactly.
Our scheme introduced in Sec.~\ref{sec:3} is based on Eq.~(\ref{eq:2.2.3}) rather than Eq.~(\ref{eq:2.2.1}).
Therefore, the skew-symmetry of Eq.~(\ref{eq:2.2.4}) is the mathematical structure
which should be preserved to maintain the energy conservation.

Here we consider the physical meaning of the skew-symmetric thermal relaxation term.
Reference~\cite{Chacon2000-1} proposed the following expression of the RFP equation:

\begin{align}
\frac{\partial H_\mathrm{s}}{\partial\mathbf{u}}f_\mathrm{s}=
-\frac{1}{8\pi}\frac{\partial}{\partial\mathbf{u}}\cdot\left(
\frac{\partial H_\mathrm{s}}{\partial\mathbf{u}}\otimes\frac{\partial H_\mathrm{s}}{\partial\mathbf{u}}
-\frac{\mathsf{I}}{2}\left|\frac{\partial H_\mathrm{s}}{\partial\mathbf{u}}\right|^2\right).\label{eq:2.2.5}
\end{align}
Equation~(\ref{eq:2.2.5}) came from similarity with the Maxwell stress tensor in electromagnetism,
and it was one of the most important structure in their paper.
Although the above formula Eq.~(\ref{eq:2.2.5}) itself is not useful for the multispecies RFP equation,
some similar terms appear in Eq.~(\ref{eq:2.2.4}).
In the Maxwell stress tensor formalism, ``$\partial H_\mathrm{s}/\partial\mathbf{u}$,
$\partial H_\mathrm{s'}/\partial\mathbf{u}$'' are kind of the electric field,
and ``$f_\mathrm{s}$, $f_\mathrm{s'}$'' are kind of the charge density of
species ``$\mathrm{s}$, $\mathrm{s'}$,'' respectively.
The primary term of Eq.~(\ref{eq:2.2.4}) means the positive work done by the Coulomb force of ``$\mathrm{s'}$''
acting on the species ``$\mathrm{s}$,'' and the secondary term means the negative counterpart.
The thermal relaxation term Eq.~(\ref{eq:2.2.4}) vanishes
when the positive and negative works equilibrate.
This is equivalent to the microscopic mechanical equilibrium between species ``$\mathrm{s}$'' and ``$\mathrm{s'}$,''
so the thermal equilibrium state is established in such a situation.
In addition, the thermal relaxation term Eq.~(\ref{eq:2.2.4}) also vanishes
in the case of like-particle collision;
this ensures that species do not gain kinetic energy from the field generated by themselves.
As a result of above discussions, the skew-symmetry can be understood as the action--reaction law
in physical sense.

\section{The structure-preserving scheme based on the discontinuous Galerkin method}\label{sec:3}
\begin{figure}
   \centering
   \includegraphics[width=0.8\textwidth]{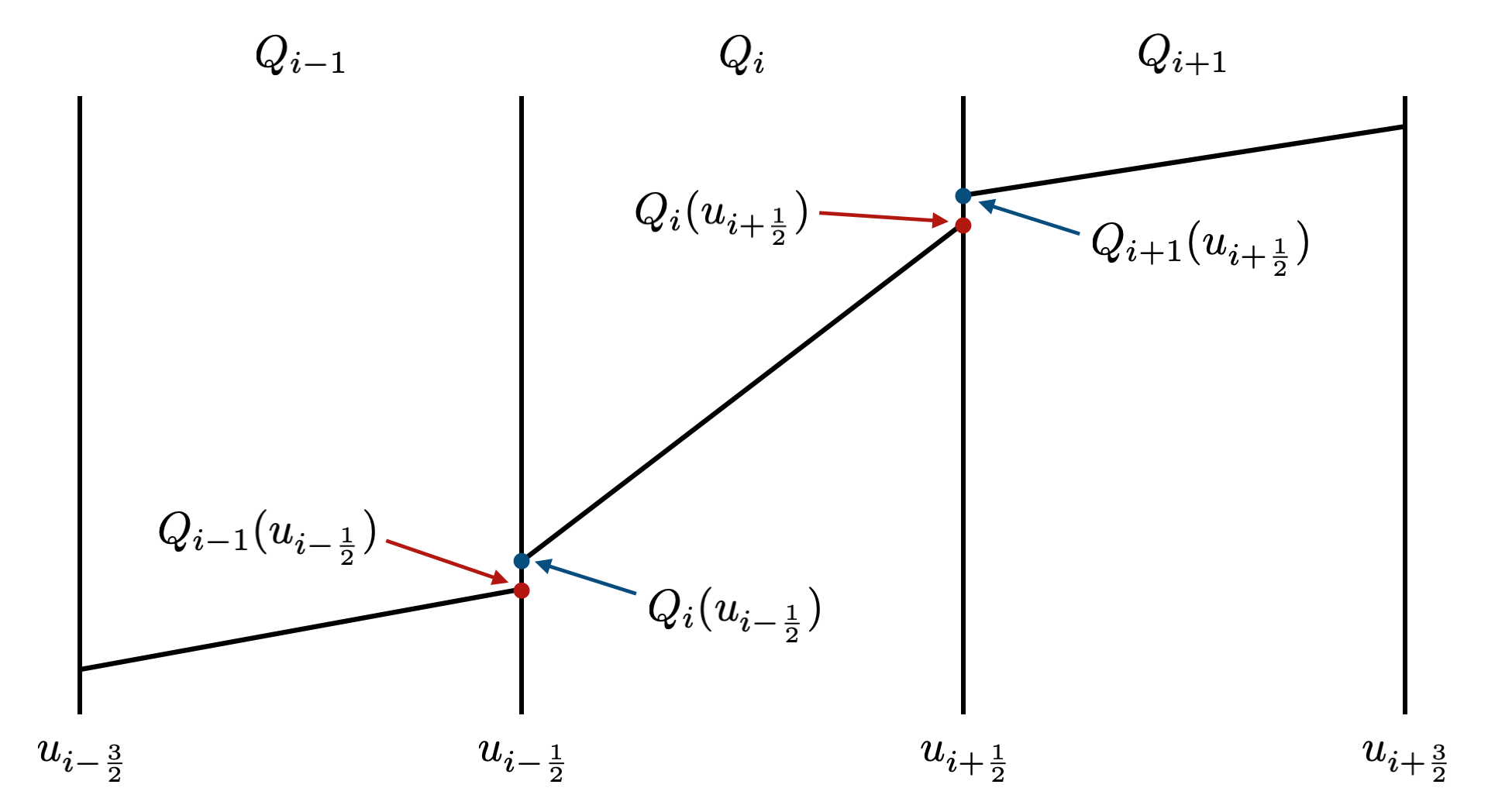}
   \caption{The indices of unknowns and grid points are defined as integer and half-integer, respectively.}
   \label{fig:1}
\end{figure}

In this section, we derive a mass-energy-conserving scheme by preserving the skew-symmetry.
The thermal relaxation term is included to the energy moment equation,
it is natural to construct a conservative scheme by the finite-element method.
The DG method gives boundary conditions through the numerical flux unlike the continuous Galerkin method,
so the conservative boundary condition is easy to be implemented in the DG method.
The DG method is designed to solve first-order PDEs,
so the second-order PDEs~(\ref{eq:2.1}) and (\ref{eq:2.2}) are
separated into a system of first-order PDEs by introducing additional unknowns as follows:

\begin{align}
\frac{\partial f_\mathrm{s}}{\partial t}=\Gamma_\mathrm{s/s'}\frac{1}{u^2}\frac{\partial}{\partial u}\left\{
u^2\left(H_\mathrm{s'}-\frac{2}{u}G_{u,\mathrm{s'}}\right)f_{u,\mathrm{s}}-
\frac{m_\mathrm{s}}{m_\mathrm{s'}}u^2 H_{u,\mathrm{s'}}f_\mathrm{s}\right\},\label{eq:3.1}\\
f_{u,\mathrm{s}}=\frac{\partial f_\mathrm{s}}{\partial u},\label{eq:3.2}\\
\frac{1}{u^2}\frac{\partial}{\partial u}\left(u^2 H_{u,\mathrm{s'}}\right)=-8\pi f_\mathrm{s'},\label{eq:3.3}\\
H_{u,\mathrm{s'}}=\frac{\partial H_\mathrm{s'}}{\partial u},\label{eq:3.4}\\
\frac{1}{u^2}\frac{\partial}{\partial u}\left(u^2 G_{u,\mathrm{s'}}\right)=H_\mathrm{s'},\label{eq:3.5}\\
G_{u,\mathrm{s'}}=\frac{\partial G_\mathrm{s'}}{\partial u},\label{eq:3.6}
\end{align}
where $f_u, H_u, G_u$ are the gradient of $f, H, G$, respectively.
As the conservation laws of mass and energy are ensured only by Eqs.~(\ref{eq:2.1.1}) and (\ref{eq:2.2.3}),
Eqs.~(\ref{eq:3.2})--(\ref{eq:3.6}) can be discretized with arbitrary combination of
numerical flux and nonlinear limiter.
Furthermore, the following redundant variable is added to use a nonlinear limiter for numerical stability:

\begin{align}
E_\mathrm{s'}=\frac{\partial (uH_\mathrm{s'})}{\partial u},\label{eq:3.7}
\end{align}
which is used in the energy equation.
The unknowns are expressed as linear combinations of basis functions:

\begin{align}
Q_i(u,t)=\sum_j Q_{i,j}(t) \phi_{i,j}(u),\label{eq:3.8}
\end{align}
where $Q$ represents the unknowns, i.e., $f, f_u, H, H_u, G, G_u, E$.
As shown in Fig.~\ref{fig:1}, the indices of unknowns and grid points are defined
as integer and half-integer, respectively.
The number of basis functions must not be less than that of the conservation laws,
so the Legendre polynomials of zeroth- and first-order, i.e. $\phi_{i,0}=1, \phi_{i,1}=u-(u_{i+\frac12}+u_{i-\frac12})/2$,
are employed in this paper.

\subsection{Discretization of Eqs.~(\ref{eq:3.3})--(\ref{eq:3.6})}\label{sec:3.1}
The types of Eqs.~(\ref{eq:3.3}) and (\ref{eq:3.4}) are identical to those of
Eqs.~(\ref{eq:3.5}) and (\ref{eq:3.6}) respectively, so only former PDEs
are discussed here.
The weak-formulation of Eq.~(\ref{eq:3.3}) is described as follows:

\begin{align}
\left[\psi_{i,k} u^2 H_{u,\mathrm{s'}}\right]_{u^-_{i-\frac12}}^{u^-_{i+\frac12}}
-\int_{u_{i-\frac12}}^{u_{i+\frac12}}\frac{\mathrm{d}\psi_{i,k}}{\mathrm{d}u}u^2H_{u,\mathrm{s'},i}\mathrm{d}u
=-8\pi\int_{u_{i-\frac12}}^{u_{i+\frac12}}\psi_{i,k}u^2f_{\mathrm{s'},i}\mathrm{d}u,\label{eq:3.1.1}
\end{align}
where $\psi$ is the test function chosen as $\psi_{i,0}=1, \psi_{i,1}=1/u$,
and the numerical flux is determined by the left-sided limit:

\begin{align}
\left[Q\right]_{u^-_{i-\frac12}}^{u^-_{i+\frac12}}\equiv Q_i(u_{i+\frac12})-Q_{i-1}(u_{i-\frac12}),\label{eq:3.1.2}
\end{align}
where the superscript of $u_{i+\frac12}^-$ means the left-sided value of $u_{i+\frac12}$.
Equation~(\ref{eq:3.4}) is discretized as follows:

\begin{align}
\int_{u_{i-\frac12}}^{u_{i+\frac12}}\psi_{i,k}H_{u,\mathrm{s'},i}\mathrm{d}u
=\left[\psi_{i,k}H_{\mathrm{s'}}\right]_{u^+_{i-\frac12}}^{u^+_{i+\frac12}}
-\int_{u_{i-\frac12}}^{u_{i+\frac12}}\frac{\mathrm{d}\psi_{i,k}}{\mathrm{d}u}H_{\mathrm{s'},i}\mathrm{d}u.\label{eq:3.1.3}
\end{align}
Test function are the same with the basis functions here, i.e., $\psi_{i,k}=\phi_{i,k}$,
and the numerical flux is the right-sided one:

\begin{align}
\left[Q\right]_{u^+_{i-\frac12}}^{u^+_{i+\frac12}}\equiv Q_{i+1}(u_{i+\frac12})-Q_i(u_{i-\frac12}),\label{eq:3.1.4}
\end{align}
where the superscript of $u_{i+\frac12}^+$ means the right-sided value of $u_{i+\frac12}$.
In this formulation, the numerical flux at outermost boundary is not decidable,
so the following farfield boundary condition should be enforced:

\begin{align}
H_\mathrm{s'}(u^+_{i_\mathrm{max}+\frac12})=\frac{8\pi}{u_{i_\mathrm{max}+\frac12}}
\int_0^{u_{i_\mathrm{max}+\frac12}}
f(u')u'^2\mathrm{d}u',\label{eq:3.1.5}
\end{align}
where $i_\mathrm{max}$ is the number of cells in the computational domain.
The distribution function is assumed to be zero outside the domain.
Likewise, the farfield boundary condition for Eq.~(\ref{eq:3.6}) is as follows:

\begin{align}
G_\mathrm{s'}(u^+_{i_\mathrm{max}+\frac12})=2\pi\int_0^{u_{i_\mathrm{max}+\frac12}}
f(u')u'^2\left(2u_{i_\mathrm{max}+\frac12}+\frac{2u'^2}
{3u_{i_\mathrm{max}+\frac12}}\right)\mathrm{d}u'.\label{eq:3.1.6}
\end{align}

\subsection{Discretization of Eqs.~(\ref{eq:3.1}), (\ref{eq:3.2}), and (\ref{eq:3.7})}\label{sec:3.2}
Equations~(\ref{eq:3.2}) and (\ref{eq:3.7}) are discretized by almost the same way;
the right-sided numerical flux is employed here:

\begin{align}
\int_{u_{i-\frac12}}^{u_{i+\frac12}}\psi_{i,k}f_{u,\mathrm{s},i}\mathrm{d}u=
\left[\psi_{i,k}f_{\mathrm{s}}\right]_{u^+_{i-\frac12}}^{u^+_{i+\frac12}}
-\int_{u_{i-\frac12}}^{u_{i+\frac12}}\frac{\mathrm{d}\psi_{i,k}}
{\mathrm{d}u}f_{\mathrm{s},i}\mathrm{d}u,\label{eq:3.2.1}\\
\int_{u_{i-\frac12}}^{u_{i+\frac12}}\psi_{i,k}E_{\mathrm{s'},i}\mathrm{d}u=
\left[\psi_{i,k}uH_{\mathrm{s'}}\right]_{u^+_{i-\frac12}}^{u^+_{i+\frac12}}
-\int_{u_{i-\frac12}}^{u_{i+\frac12}}\frac{\mathrm{d}\psi_{i,k}}
{\mathrm{d}u}uH_{\mathrm{s'},i}\mathrm{d}u,\label{eq:3.2.2}
\end{align}
where the boundary conditions for Eq.~(\ref{eq:3.2.1}) is $f_\mathrm{s}(u^+_{n+\frac12})=0$,
and the far-field condition (\ref{eq:3.1.5}) for Eq.~(\ref{eq:3.2.2}).
As mentioned above, a minmod function is used as the nonlinear limiter:

\begin{align}
\widetilde{E}_{\mathrm{s'},i,0}=E_{\mathrm{s'},i,0},\label{eq:3.2.3}\\
\widetilde{E}_{\mathrm{s'},i,1}=\mathrm{minmod}\left(E_{\mathrm{s'},i,1},
\frac{E_{\mathrm{s'},i+1,0}-E_{\mathrm{s'},i,0}}{u_{i+\frac12}-u_{i-\frac12}},
\frac{E_{\mathrm{s'},i,0}-E_{\mathrm{s'},i-1,0}}{u_{i+\frac12}-u_{i-\frac12}}
\right),\label{eq:3.2.4}\\
\mathrm{minmod}(a,b,c)\equiv\begin{cases}
\min(a,b,c)\quad\mathrm{if}\quad a,b,c>0,\\
\max(a,b,c)\quad\mathrm{if}\quad a,b,c<0,\\
0\quad\mathrm{otherwise},
\end{cases}\label{eq:3.2.5}
\end{align}
where $\widetilde{E}$ is the result of minmod operation.
Finally, Eq.~(\ref{eq:3.1}) is discretized as follows to preserve the mass-energy conservation simultaneously.

\begin{align}
\int_{u_{i-\frac12}}^{u_{i+\frac12}} \frac{\partial f_{\mathrm{s},i}}{\partial t}m_\mathrm{s}u^2\mathrm{d}u
&=m_\mathrm{s}^2\Gamma_\mathrm{s/s'}\left[\frac{1}{m_\mathrm{s}}u^2\frac{\partial^2 G_\mathrm{s'}}{\partial u^2}
f_{u,\mathrm{s}}\right]_{u^-_{i-\frac12}}^{u^-_{i+\frac12}}
-m_\mathrm{s}^2\Gamma_\mathrm{s,s'}\left[\frac{1}{m_\mathrm{s'}}
u^2 H_{u,\mathrm{s'}}f_\mathrm{s}\right]_{u^+_{i-\frac12}}^{u^+_{i+\frac12}},\label{eq:3.2.6}\\
\int_{u_{i-\frac12}}^{u_{i+\frac12}} \frac{\partial f_{\mathrm{s},i}}{\partial t}m_\mathrm{s}\frac12u^4\mathrm{d}u
&=m_\mathrm{s}^2\Gamma_\mathrm{s/s'}\left[\frac{1}{2m_\mathrm{s}}u^4\frac{\partial^2 G_\mathrm{s'}}{\partial u^2}
f_{u,\mathrm{s}}\right]_{u^-_{i-\frac12}}^{u^-_{i+\frac12}}
-m_\mathrm{s}^2\Gamma_\mathrm{s/s'}\left[\frac{1}{2m_\mathrm{s'}}u^4
H_{u,\mathrm{s'}}f_\mathrm{s}\right]_{u^+_{i-\frac12}}^{u^+_{i+\frac12}}\notag\\
&-m_\mathrm{s}^2\Gamma_\mathrm{s/s'}\left[\frac{1}{m_\mathrm{s}}u^3
\frac{\partial^2 G_\mathrm{s'}}{\partial u^2}f_\mathrm{s}\right]_{u^-_{i-\frac12}}^{u^-_{i+\frac12}}
-\frac{m_\mathrm{s}^2\Gamma_\mathrm{s/s'}}{8\pi m_\mathrm{s}}
\left[u^2\widetilde{E}_\mathrm{s'}
H_{u,\mathrm{s}}\right]_{u^+_{i-\frac12}}^{u^+_{i+\frac12}}\notag\\
&+m_\mathrm{s}^2\Gamma_\mathrm{s/s'}\int_{u_{i-\frac12}}^{u_{i+\frac12}}
\left(\frac{u^3f_{\mathrm{s},i}}{m_\mathrm{s'}} H_{u,\mathrm{s'},i}
-\frac{u^3f_{\mathrm{s'},i}}{m_\mathrm{s}}H_{u,\mathrm{s},i}\right)\mathrm{d}u,\label{eq:3.2.7}\\
\frac{\partial ^2 G_\mathrm{s'}}{\partial u^2}&=H_\mathrm{s'}-\frac{2}{v}G_{u,\mathrm{s'}}.\label{eq:3.2.8}
\end{align}
In this study, the minmod limiter \cite{Roe1986} is only used on $E_\mathrm{s'}$,
so monotonicity of the distribution function is not maintained exactly.
If the total variation diminishing (TVD) limiters were used on the convective terms
of Eqs.~(\ref{eq:3.2.6}) and (\ref{eq:3.2.7}),
the spatial accuracy would be degraded to the first-order in exchange for numerical stability
because the convective terms had continuous extrema.
The proposed scheme seems to be stable in the numerical experiments performed later owing to physical dissipation.
If the physical dissipation terms are not enough to mitigate numerical instabilities,
one can use additional nonlinear limiters such as weighted essentially non-oscillatory (WENO) scheme.
Although many WENO reconstruction schemes are proposed for the DG method \cite{Qiu2003,Qiu2005},
the WENO reconstruction would destroy the conservative properties of Eqs.~(\ref{eq:3.2.6}) and (\ref{eq:3.2.7}).
Therefore, the slope limiter like the finite-volume approach \cite{Harten1987,Shu1989}
is the only solution to enhance the numerical stability.

\section{Experiments}\label{sec:4}

\begin{figure}
\begin{minipage}{0.5\textwidth}
\centering
\includegraphics[width=\textwidth]{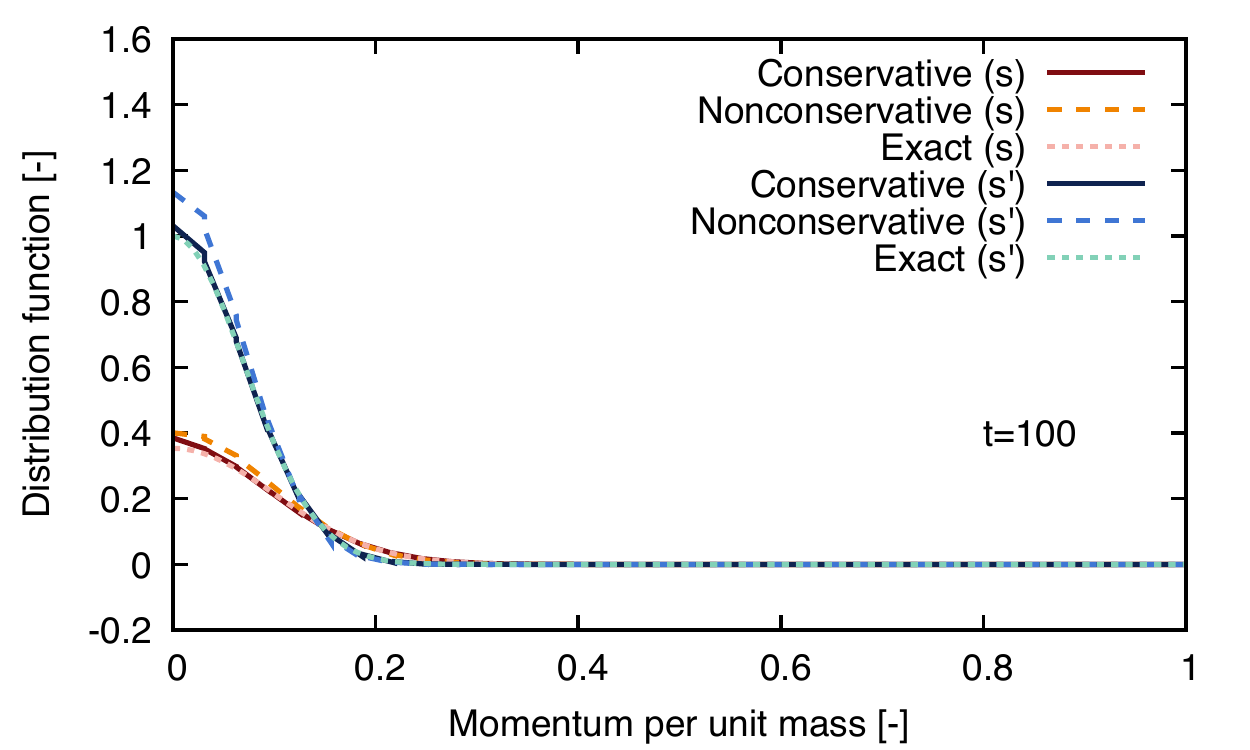}
\subcaption{\label{fig:4.1.1a} $t=100$.}
\end{minipage}
\begin{minipage}{0.5\textwidth}
\centering
\includegraphics[width=\textwidth]{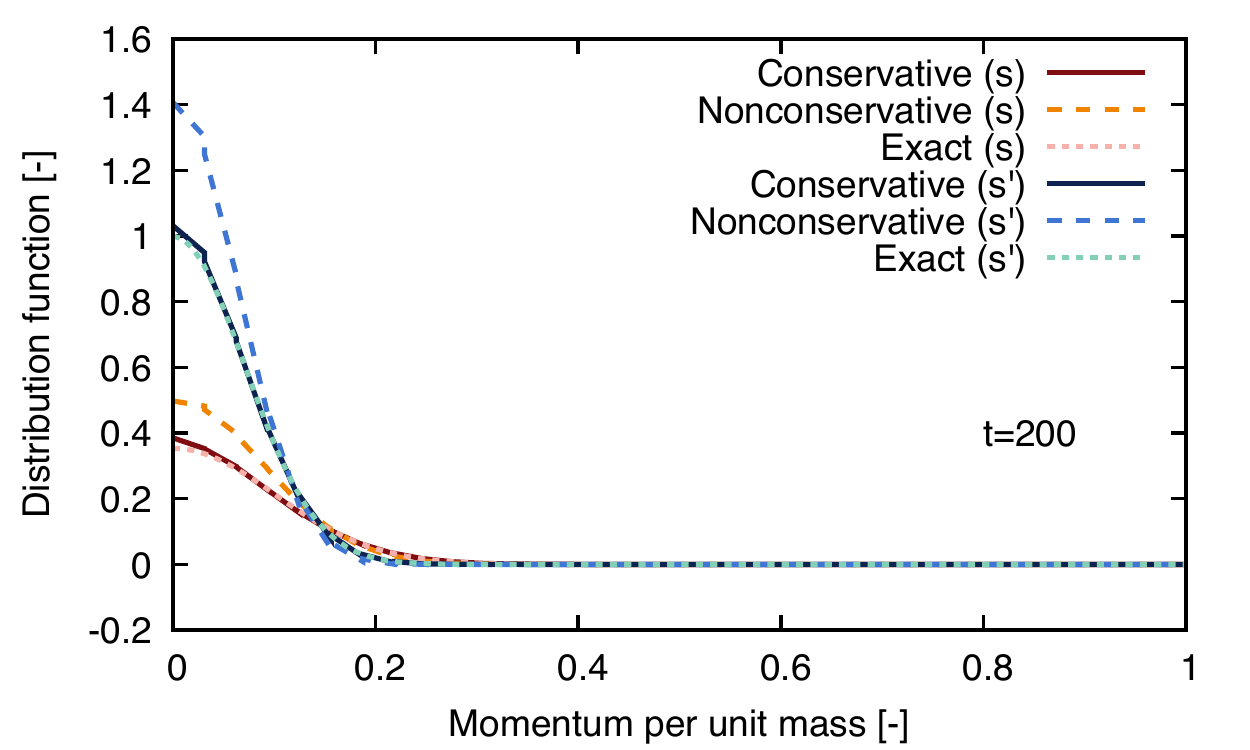}
\subcaption{\label{fig:4.1.1b} $t=200$.}
\end{minipage}
\caption{\label{fig:4.1.1} Snapshots of the distribution functions calculated by the examined schemes
in the equilibrium preservation problem.}
\end{figure}

\begin{figure}
\begin{minipage}{0.5\textwidth}
\centering
\includegraphics[width=\textwidth]{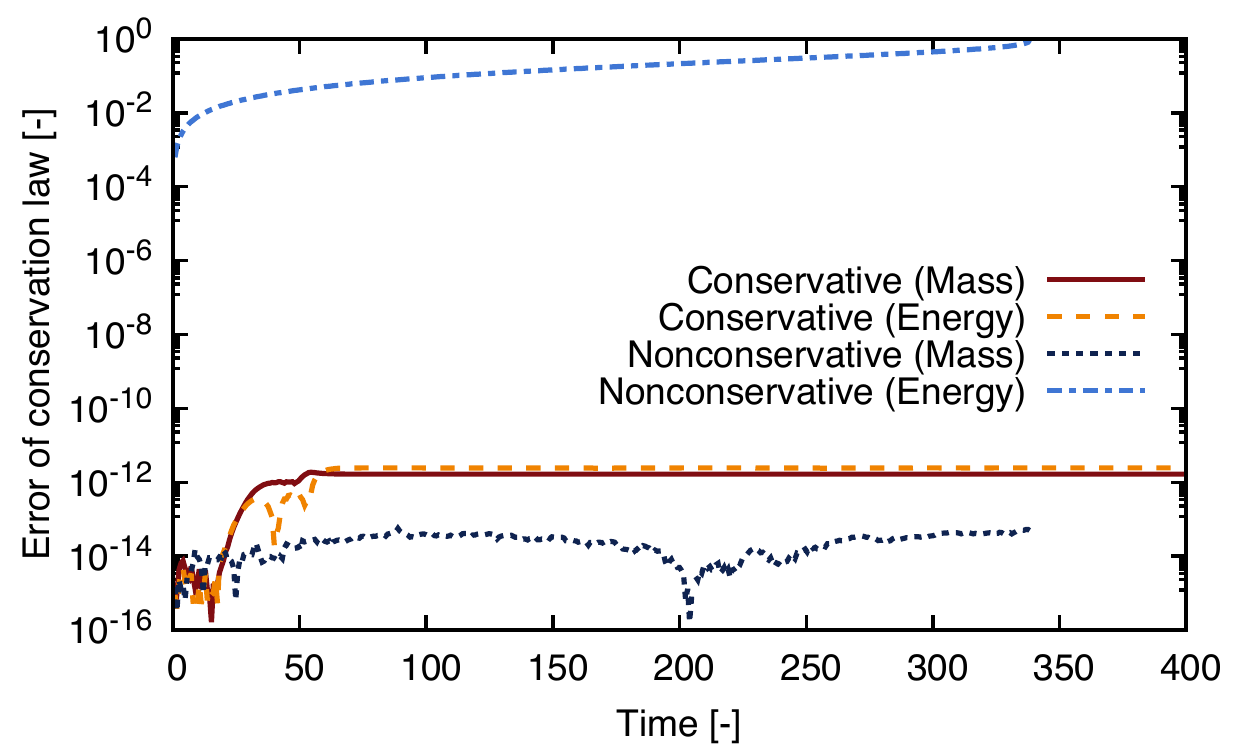}
\caption{\label{fig:4.1.2} Errors of the conservation laws in the equilibrium preservation problem.}
\end{minipage}
\begin{minipage}{0.5\textwidth}
\centering
\includegraphics[width=\textwidth]{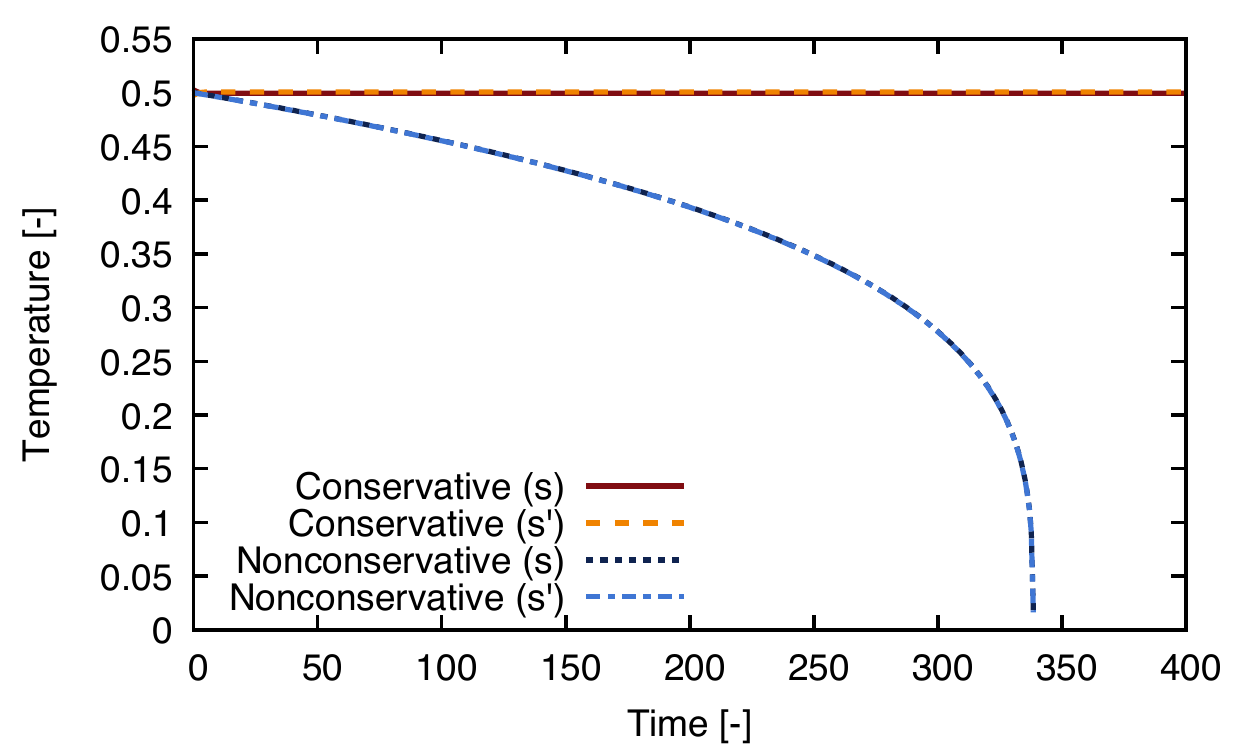}
\caption{\label{fig:4.1.3} Temperature of each species in the equilibrium preservation problem.}
\end{minipage}
\end{figure}

In this section, the conservative scheme based on Eq.~(\ref{eq:2.2.3}) is compared with
a nonconservative discretization based on Eq.~(\ref{eq:2.2.1}).
The numerical experiments take into account both like-particle and unlike-particle collision.
First, the conservative and nonconservative schemes are examined
whether an analytic thermal equilibrium can be maintained against discretization errors.
The initial conditions are set as follows:

\begin{align}
m_\mathrm{s}=1,\quad m_\mathrm{s'}=2,\quad
m_\mathrm{s}^2\Gamma_\mathrm{s/s'}=m_\mathrm{s'}^2\Gamma_\mathrm{s'/s}=1,\label{eq:4.1.1}\\
f_{0,*}=0.001\left(50m_*\right)^{3/2}\exp\left(-50m_*u^2\right),\label{eq:4.1.2}
\end{align}
where $*=\mathrm{s,s'}$, and $f_{0}$ is the initial distribution.
Equation~(\ref{eq:4.1.2}) represents the Maxwell distribution with the same temperature for different species,
so the distribution function should not evolve since it has already been thermalized.
The distribution function at $t=100$ and $t=200$ is shown in Fig.~\ref{fig:4.1.1}.
The nonconservative scheme clearly fails to preserve the equilibrium.
In contrast, the conservative scheme maintains the initial distribution well
although the numerical solution includes small truncation errors compared to the exact solution.
Figure~\ref{fig:4.1.2} displays time history of the global conservation errors.
The conservative scheme maintains the conservation laws only with the round-off errors.
On the other hand, the energy conservation is violated by the nonconservative scheme,
and the simulation crashes at $t\sim 340$ when error of the energy conservation reaches $100\%$.
Time evolution of temperature is also shown in Fig.~\ref{fig:4.1.3}.
The nonconservative scheme suffers an accelerating numerical cooling,
while the conservative scheme maintains the initial temperature well.
Table~\ref{tab:1} shows effective order of accuracy for the conservative scheme.
The time stepping is determined by a scaling of $\Delta t \propto {i_\mathrm{max}}^{-2}$
since time integration is performed by the Euler explicit method.
The order of accuracy has a good agreement with the formal accuracy when the distribution function is resolved well.

\begin{table}[h]
\centering
\caption{\label{tab:1} Effective order of accuracy in the equilibrium preservation problem.}
\begin{tabular}{ccccc}\hline
$i_\mathrm{max}$ & $\Delta t$ & $T$ & $T'$ & Order of Accuracy\\ \hline
32  & $8\times10^{-5}$ &  0.4994714090624359 & 0.5005285909351064 & \\
45  & $4.045\times10^{-5}$ &  0.4995837291042263 & 0.5004162708957715 & 0.700675\\
64  & $2\times10^{-5}$ &  0.4997560759628925 & 0.5002439240373561 & 1.51746\\
90  & $1.011\times10^{-5}$ &  0.4998677212880675 & 0.5001322787119834 & 1.79495\\
128  & $0.5\times10^{-5}$ & 0.4999319735574804 & 0.5000680264473013 & 1.88806\\
180  & $0.2528\times10^{-5}$ & 0.4999649603118326 & 0.5000350397070178 & 1.94592\\ \hline
Exact  &  & 0.5 & 0.5 & 2\\ \hline
\end{tabular}
\end{table}

\begin{figure}
\begin{minipage}{0.5\textwidth}
\centering
\includegraphics[width=\textwidth]{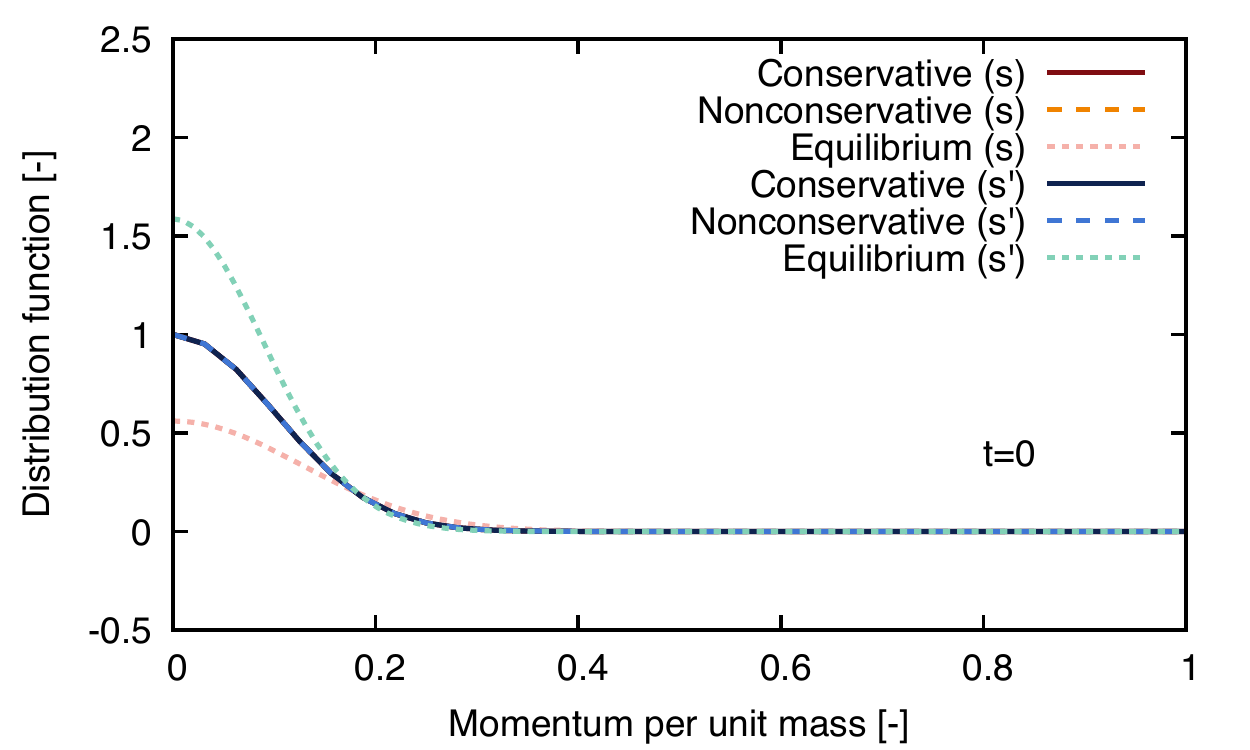}
\subcaption{\label{fig:4.2.1a} $t=0$.}
\end{minipage}
\begin{minipage}{0.5\textwidth}
\centering
\includegraphics[width=\textwidth]{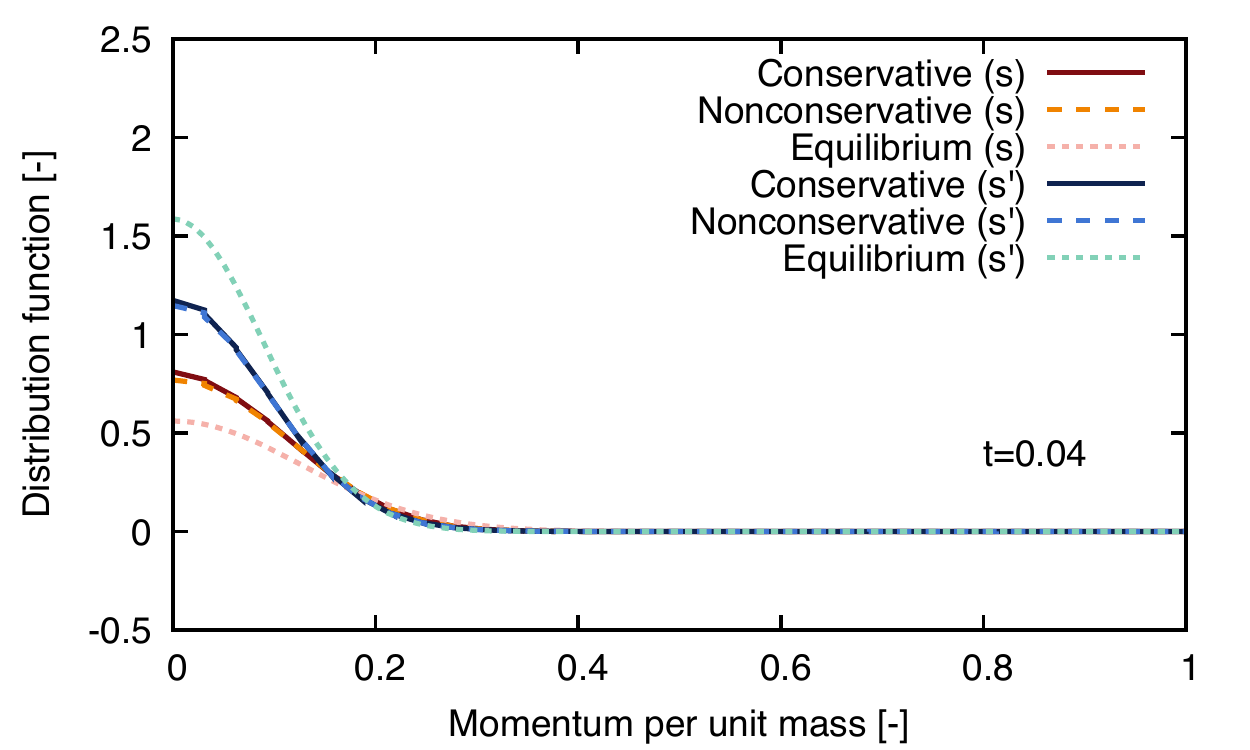}
\subcaption{\label{fig:4.2.1b} $t=0.04$.}
\end{minipage}
\begin{minipage}{0.5\textwidth}
\centering
\includegraphics[width=\textwidth]{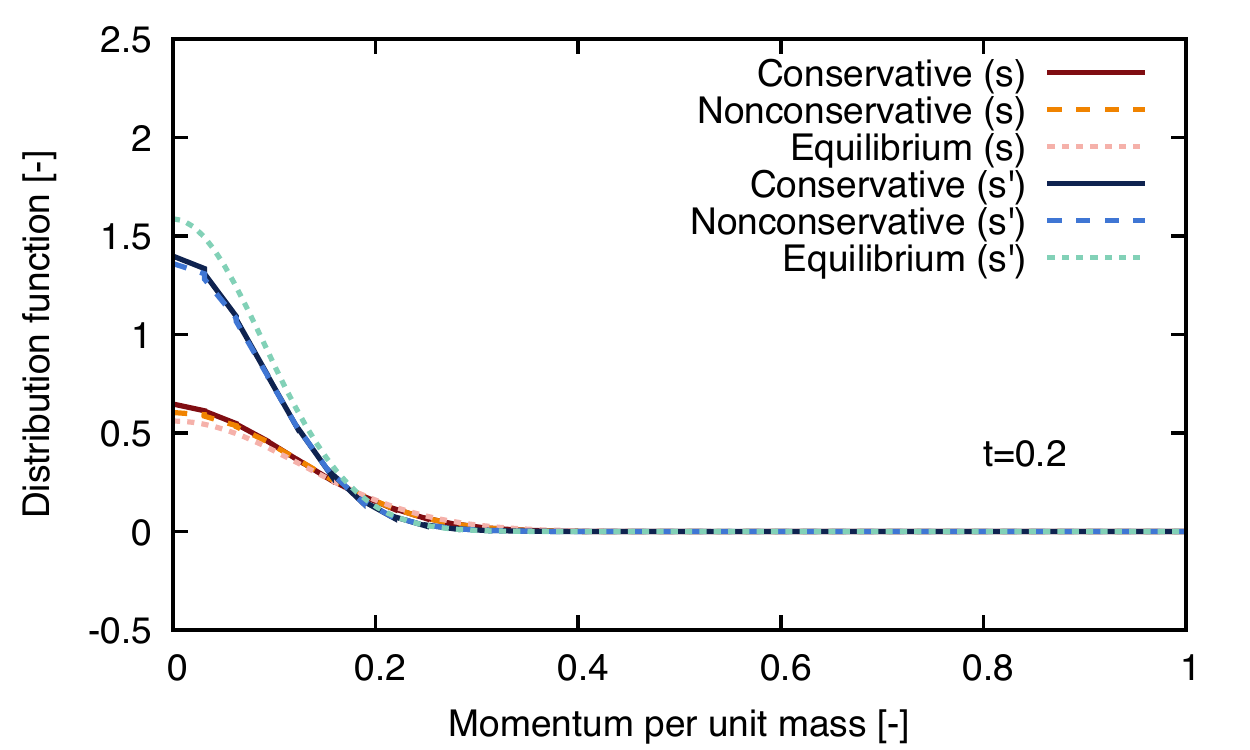}
\subcaption{\label{fig:4.2.1c} $t=0.2$.}
\end{minipage}
\begin{minipage}{0.5\textwidth}
\centering
\includegraphics[width=\textwidth]{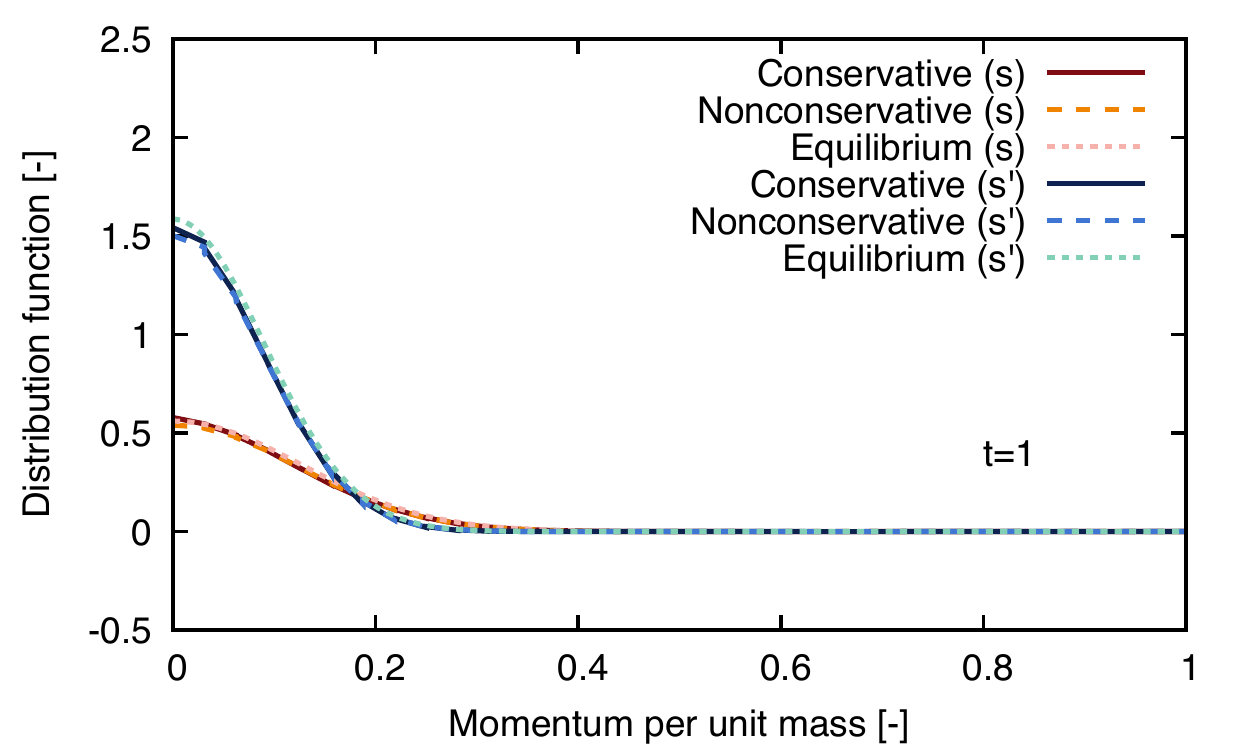}
\subcaption{\label{fig:4.2.1d} $t=1$.}
\end{minipage}
\begin{minipage}{0.5\textwidth}
\centering
\includegraphics[width=\textwidth]{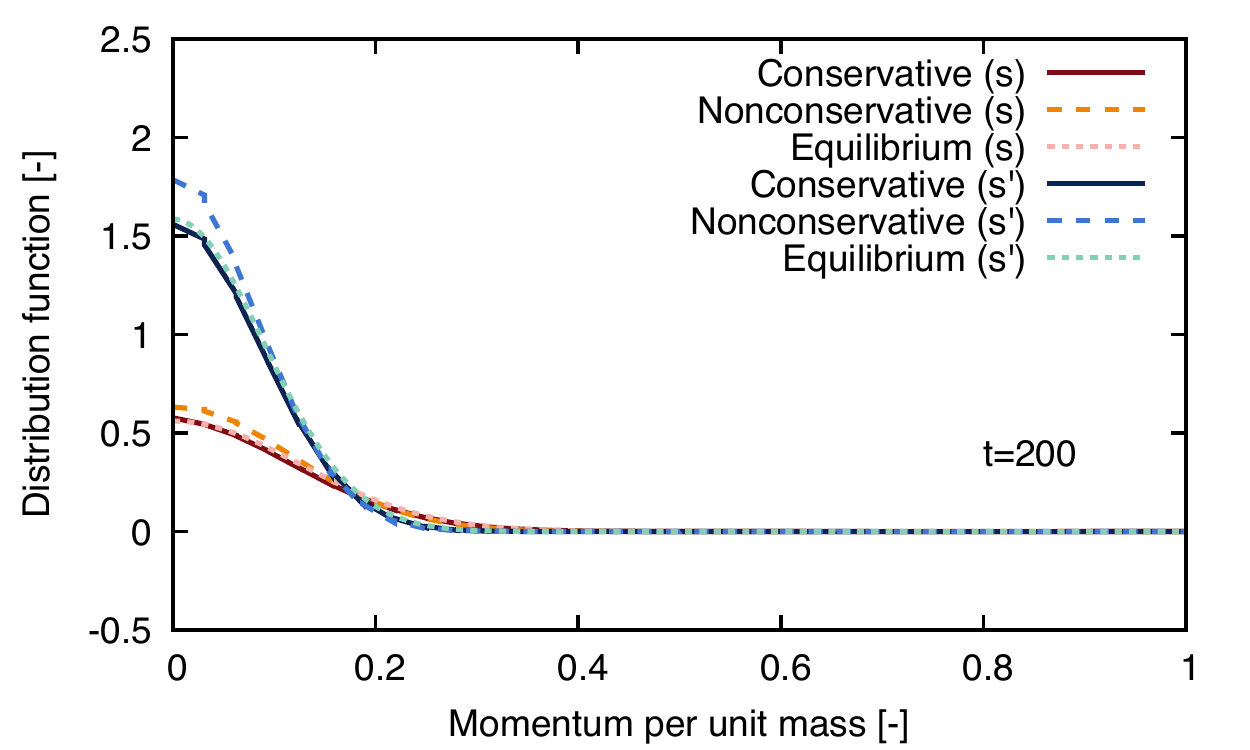}
\subcaption{\label{fig:4.2.1e} $t=200$.}
\end{minipage}
\begin{minipage}{0.5\textwidth}
\centering
\includegraphics[width=\textwidth]{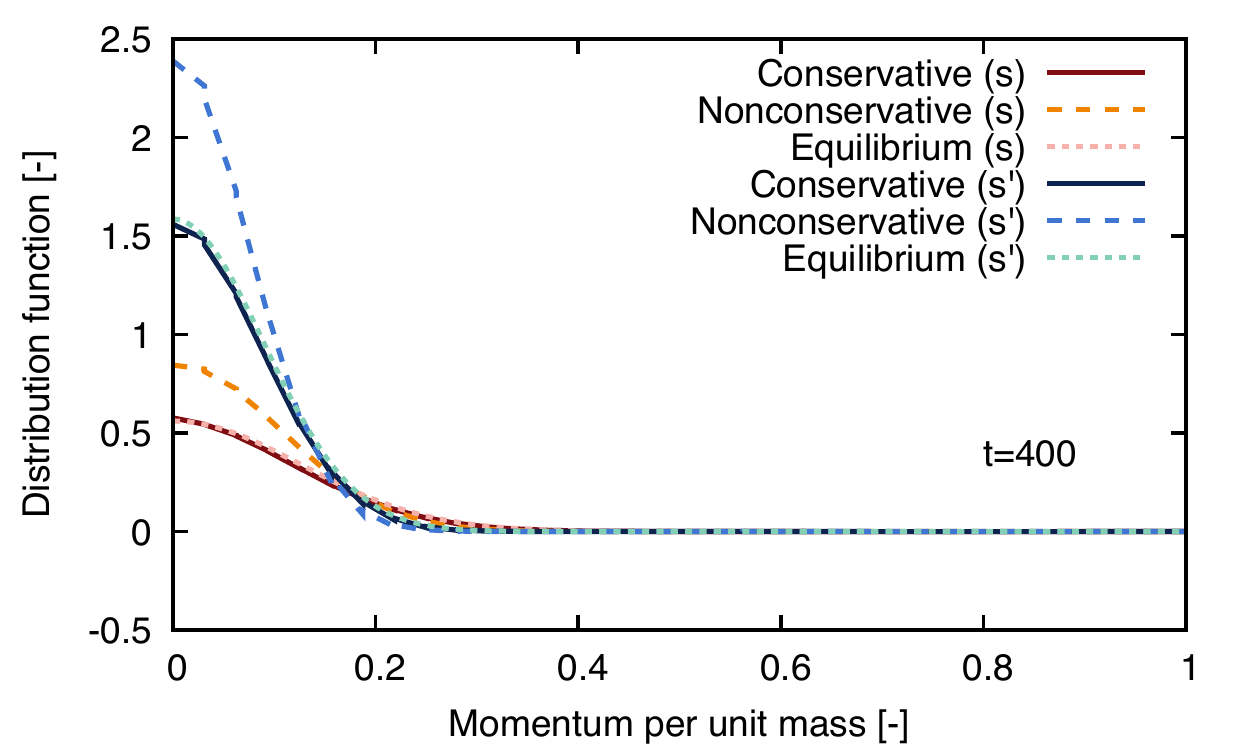}
\subcaption{\label{fig:4.2.1f} $t=400$.}
\end{minipage}
\caption{\label{fig:4.2.1} Snapshots of the distribution functions calculated by the examined schemes
in the thermal equilibration problem.}
\end{figure}

\begin{figure}
\begin{minipage}{0.5\textwidth}
\centering
\includegraphics[width=\textwidth]{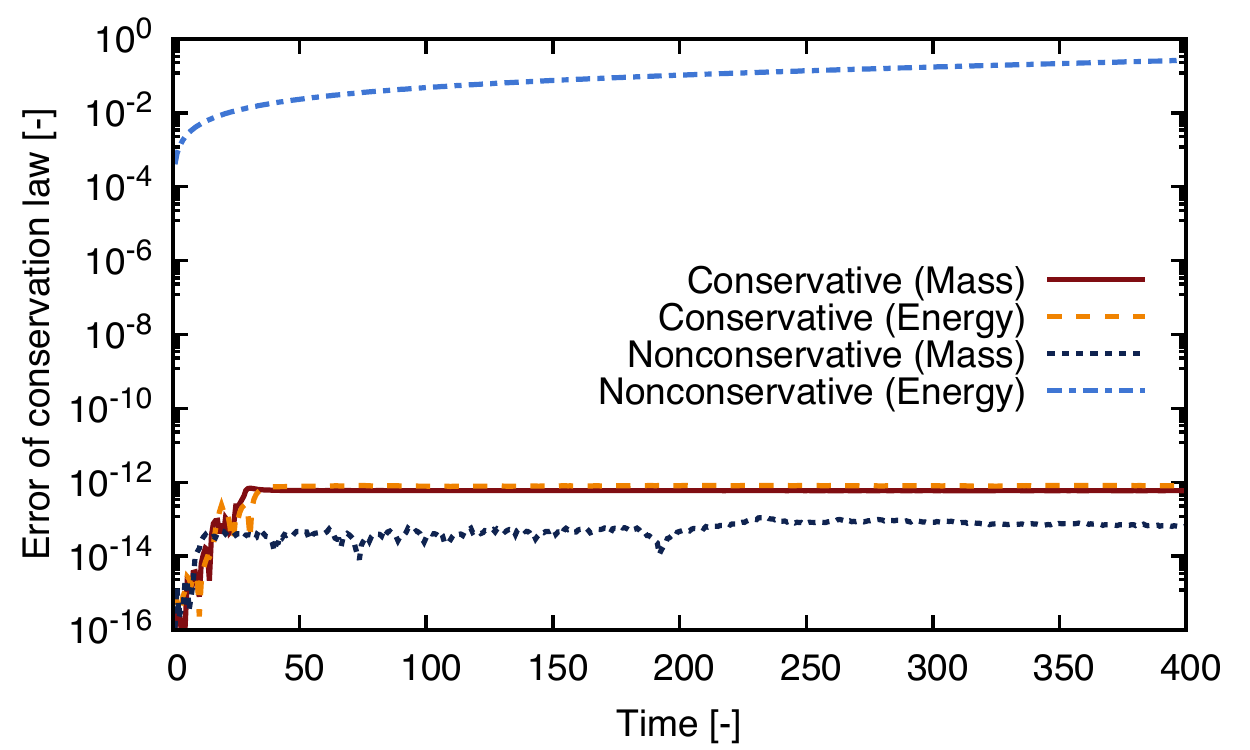}
\caption{\label{fig:4.2.2} Errors of the conservation laws in the thermal equilibration problem.}
\end{minipage}
\begin{minipage}{0.5\textwidth}
\centering
\includegraphics[width=\textwidth]{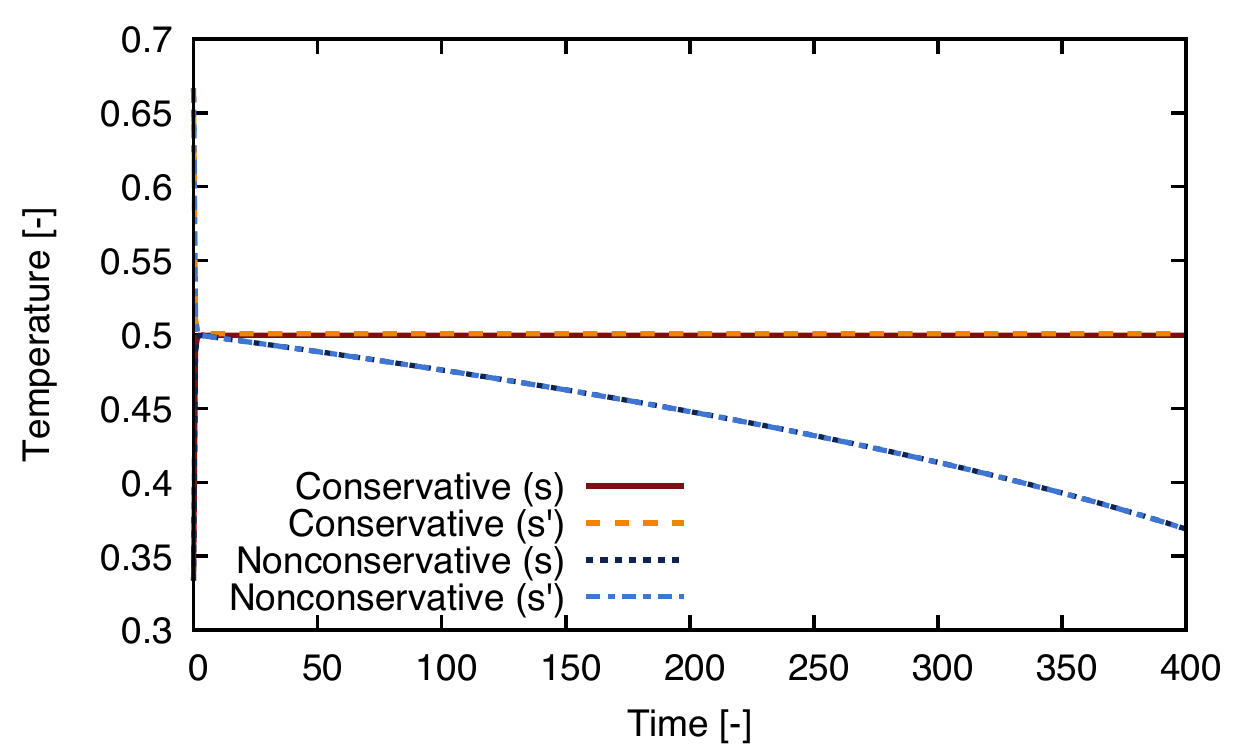}
\caption{\label{fig:4.2.3} Temperature of each species in the thermal equilibration problem.}
\end{minipage}
\end{figure}

Another example is a nonequilibrium problem in which the distribution functions are initialized with different temperatures.

\begin{align}
m_\mathrm{s}=1,\quad m_\mathrm{s'}=2,\quad
m_\mathrm{s}^2\Gamma_\mathrm{s/s'}=m_\mathrm{s'}^2\Gamma_\mathrm{s'/s}=1,\label{eq:4.2.1}\\
f_{0,*}=\exp\left(-50u^2\right).\label{eq:4.2.2}
\end{align}
The initial distribution does not include the information of mass,
so this initial setup does not mean the equilibrium state unlike the previous problem.
Figure~\ref{fig:4.2.1} shows time evolution of the distribution function.
Although the numerical solution of both schemes agree well until the equilibration ($t\lesssim 1$),
the nonconservative scheme cannot maintain the equilibrium after that
as shown in Figs.~\ref{fig:4.2.1e} and \ref{fig:4.2.1f}.
Figures~\ref{fig:4.2.2} and \ref{fig:4.2.3} are the errors of conservation laws
and time evolution of the temperature, respectively.
Both results show the same trend with the static numerical experiment.

\section{Conclusions}\label{sec:5}
In this paper, we demonstrated that a mass-energy-conserving scheme for the isotropic Rosenbluth--Fokker--Planck equation
can be composed by preserving mathematical/physical structure of the system.
The key point to realize conservative multispecies simulation is that the volume integral of the energy moment equation
is transformed into a skew-symmetric form by integration-by-parts.
Although our previous works on structure-preserving kinetic schemes depend on linearity of the central difference scheme,
the present scheme can accept nonlinear upwind schemes
which are mandatory for numerical stability of convection terms.
The conservative scheme is compared with a conventional scheme
through some numerical experiments.
Although the mass conservation is preserved by both schemes,
the conservative scheme is the only one which can conserves the total energy strictly.
The conservative scheme also reproduces equilibration process accurately
while the nonconservative simulation experiences a fatal crash
when the total energy becomes negative due to numerical cooling.
Therefore, conservative Fokker--Planck simulation is also demonstrated for the Rosenbluth formulation.
The derivation of conservation laws are done by an analytic approach in this paper,
so a relativistic extension would be performed by the same strategy.

\section*{Acknowledgment}
This work was partially supported by KAKENHI (19K21038, 20K14449).


\end{document}